\documentclass[useAMS,usenatbib]{mn2e}
\usepackage{graphicx,color,epsfig,natbibmnfix}

\newcommand{\be}{\begin{equation}}
\newcommand{\ee}{\end{equation}}

\newcommand{\apj}{ApJ}

\newcommand{\mnras}{MNRAS}
\newcommand{\aap}{A\&A}

\newcommand{\apjl}{ApJL}

\def\ltsima{$\; \buildrel < \over \sim \;$}
\def\simlt{\lower.5ex\hbox{\ltsima}}
\def\gtsima{$\; \buildrel > \over \sim \;$}
\def\simgt{\lower.5ex\hbox{\gtsima}}

\def\msun{{\,{\rm M}_\odot}}

\def\lsun{{\,L_\odot}}
\def\del#1{{}}
\title[Radiative feedback from protoplanets]{Radiative feedback from
  protoplanets in self-gravitating protoplanetary discs}

\author[Sergei Nayakshin and Seung-Hoon Cha]{Sergei Nayakshin$^1$ and Seung-Hoon
  Cha$^{1,2}$\\ $^1$ Department of Physics \& Astronomy, University of Leicester,
  Leicester, LE1 7RH, UK\\ 
$^2$ Department of Physics \& Astronomy, Texas A\&M University-Commerce,
  P.O. Box 3011, Commerce, Texas, 75429, USA
}
\begin{document}

\date{Received}

\pagerange{\pageref{firstpage}--\pageref{lastpage}} \pubyear{2011}

\maketitle

\label{firstpage}

\begin{abstract}
It is well known that massive protoplanetary disc are gravitationally unstable
beyond tens of AU from their parent star. The eventual fate of the
self-gravitating gas clumps born in the disc is currently not understood,
although the range of uncertainty is well known. If clumps migrate inward
rapidly, they are tidally disrupted, which may leave behind giant or
terrestrial like planets. On the other hand, if clumps migrate less rapidly,
they tend to accrete gas, becoming proto brown dwarfs or low mass companions
to the parent star.

Here we argue that radiative feedback of contracting clumps
(protoplanets) on their discs is an important effect that has been
overlooked in previous calculations. We show analytically that
temperature in clump's vicinity may be high enough to support a
quasi-static atmosphere if the clump mass is below a critical value,
$M_{\rm cr} \sim 6$ Jupiter masses ($M_J$). This may arrest further
gas accretion onto the clump and thus promote formation of planets
rather than low mass companions.

We use numerical simulations to evaluate these analytical conclusions
by studying migration and accretion of gas clumps as a function of
their initial mass, $M_i$. Simulations neglecting the radiative
preheating effect show that gas clumps with mass less than $\sim 2
M_J$ migrate inward rapidly; more massive clumps result in low mass
companions. In contrast, simulations that include radiative preheating
from the clump show that clumps as massive as $8 M_J$ migrate inward
rapidly and end up tidally disrupted in the inner disc.  We conclude
that, with other parameters being equal, previous simulations
neglecting radiative feedback from self-gravitating clumps
over-estimated the population of brown dwarfs and low mass stellar
companions and under-estimated the population of planets.
\end{abstract}

%\keywords{}

\section{Introduction}\label{sect:intro}

Massive (e.g., $> 0.1 \msun$) protoplanetary discs are expected to cool
rapidly enough to fragment beyond tens to a hundred AU, depending on
parameters of the problem
\citep{Gammie01,MayerEtal04,Rice05,Rafikov05,DurisenEtal07,SW08,Meru10}.  The
exact fate of these clumps is however not well known because of the physical
and numerical uncertainties of the calculations \citep[e.g.,][]{SW09}, and
because of a huge parameter space (e.g., disc opacity, dust physics more
generally, angular momentum of the incoming material, mass deposition rate
into the disc, external illumination, etc.).

A subset of authors, e.g., \cite{VB06,VB10}, \cite{BoleyEtal10},
\cite{ChaNayakshin11a}, \cite{MachidaEtal11} find repeated clump formation
followed by inward clump migration and tidal destruction in the inner disc.
This process inspired a new ``Tidal Downsizing'' scenario for planet formation
\citep{BoleyEtal10,Nayakshin10c} that can {\em potentially} explain formation
of all types of planets, although much work is to be done to rigorously test
and develop this model further.

Two further publications, \cite{MichaelEtal11} and
\cite{BaruteauEtal11} presented simulations of marginally stable
self-gravitating discs with a single planet inserted into the disc and
treated as a point mass (which disallows further gas accretion onto
the planet), and found that the planets did migrate inward on the time
scale of just a few orbits.

On the other hand, a number of other authors did not find such a strong clump
migration. For example, disc simulations by \cite{SW08,SW09b} show formation
of many more brown dwarfs than planet-mass objects, and these do not appear to
migrate inward in the discs that much. Note that earlier $\beta$-prescription
cooling simulations, such as \cite{Rice05}, cannot be useful here because
these simulations typically stall due to increasingly small timesteps needed
to integrate dense clump evolution when one or more clumps are
formed. Finally, a recent study by \cite{ZhuEtal12a} find that the outcome of
clump formation is not a single outcome process, with 6 of their 13 clumps
migrating in and crossing the inner boundary, 4 migrating and being tidally
destroyed before reaching the inner boundary, and 3 becoming brown dwarfs and
stalling their migration in the outer disc as they open deep
gaps. \cite{ZhuEtal12a} suggested that it is a competition between clump
migration and accretion onto the clump that appears to decide which of these
outcomes takes place.

In this paper we point out yet another physical complication than will
have to be modelled robustly enough before the community arrives at a
firm conclusion about the non-linear outcome of the gravitational disc
instability. In particular, all of the gravitationally unstable disc
simulations performed so far employed a prescription of some kind to
approximate radiative transfer out of the disc. None of these looked
into how the energy release within the gas clump (protoplanet) may
affect the surrounding gas in the disc. To see that this
``preheating'' effect may be significant, consider the accretion
luminosity of the {\em whole} protostellar disc at the unstable
region, $R \sim 100$~AU:
\begin{equation}
L_{\rm disc} \sim {G M_* \dot M_* \over 2R} \approx 10^{-3} \lsun {\dot
  M_*\over 10^{-6} \msun\hbox{yr}^{-1}}\;,
\label{ldisk}
\end{equation}
where $M_* \approx 1\msun$ is the mass of the protostar, and $\dot M_*$ is the
accretion rate in the disc. 

Now protoplanetary clumps of a few to 10 $M_J$ can be as bright as this or
even brighter at birth \citep[e.g.,][]{Nayakshin10a,ZhuEtal12a}. This
luminosity is therefore bound to significantly heat up the gas near the
clump's location. 

We present in \S \ref{sec:analytical} ``radiative zero atmosphere'' analytical
arguments, familiar from the physically similar situation encountered during
the growth of giant planets in the CA scenario
\citep{Mizuno80,Stevenson82}. These arguments show that the contraction
luminosity of the young protoplanet supports a hot atmosphere around the
protoplanet, preventing further accretion of gas onto it. Just as in CA
theory, we show that there should exist a critical protoplanet mass, $M_{\rm
  cr}$, above which the radiative atmosphere solution is unstable to
self-gravity and should collapse onto the protoplanet. The implications of
this argument are that previous numerical simulations that neglected radiative
feedback from the protoplanets may have over-estimated the accretion rate of
gas onto the clumps, thus over-produced low mass stellar companions and brown
dwarfs, and under-produced lower mass objects -- planets.

We run numerical simulations with and without protoplanetary feedback
to investigate these ideas more quantitatively. We set up a disc in a
marginally gravitationally stable regime and then insert a density
enhancement of a specified total mass inside of a spiral arm. We
confirm the \cite{ZhuEtal12a} findings that the fragments stall when
they accrete a significant amount of gas from the disc. Specifically,
without protoplanetary feedback, only protoplanets with initial mass
$M_p\simlt 2 M_J$ migrated inward faster than they accreted gas. These
planets migrated inward sufficiently fast to be tidally
disrupted. However protoplanets with masses $M_p$ larger than that
gained too much mass too quickly and stalled, becoming proto brown
dwarfs or low mass stellar companions rather than planets.

Simulations which take the protoplanetary feedback into account via a
prescription (see \S \ref{sec:rad_heat}) show that gas accretion is indeed
significantly slowed down, so that clumps with an initial mass of as much as
$8 M_J$ migrate rapidly and are tidally disrupted.

We conclude that the non-linear outcome of clump formation in the
outer self-gravitating disc strongly depends on the temperature
structure near the protoplanet. Future work must include the
preheating effects in order to obtain reliable results.

\section{A quasi-equilibrium atmosphere around the gas clump}\label{sec:analytical}

First of all, we make a note on terminology: it is not exactly clear when one
should call a self-gravitating gas clump ``a planet''. One may choose to
consider Hydrogen molecules dissociation and formation of a very dense
``second core'' as a sign post of planet formation \citep[e.g.,][]{SW09}. On
the other hand, such a planet may accrete more mass from the disc, becoming a
brown dwarf or even a low mass star, or, on the contrary, migrate very close
to the parent star and be unbound by tides at separations of $\sim 0.1$~AU
\citep{Nayakshin11b}. This may leave just a solid core behind or nothing at
all, if the ``planet'' was already too hot for dust sedimentation
\citep{HS08}. In this paper we cannot trace all these possible outcomes, so we
simply define a planet as any ``obvious'' dense self-gravitating gas
condensation. Therefore, a ``gas clump'' and a ``planet'' are sinonimous here.

Before we begin our numerical investigation, we appeal to the
following parallel: accretion of the gaseous envelope on the top of a
solid core immersed in a protoplanetary disc in the context of the CA
model \citep[e.g.,][]{Mizuno80}. It is well known that at a given core
luminosity, $L_c$, there exists a critical core mass, $M_{\rm
  cr}$. Cores with mass less than $M_{\rm cr}$ are surrounded by a
gaseous hydrostatic atmosphere. Atmospheres of cores with mass larger
than $M_c$ are too massive (comparable in mass to $M_{\rm cr}$) for a
hydrostatic solution to apply; they ``collapse'' onto the solid
core. This sets off an accretion phase onto the solid core and
eventually results in formation of a giant planet.

The situation we are concerned about is actually analogous to this classical
result from the CA theory, except that the role of the dense core is played by
the dense -- mainly gaseous -- protoplanet. The luminosity of the protoplanet
here is not due to accretion of planetesimals but is rather due to
contraction of the protoplanet. 

We follow the analytical argument first presented by \cite{Stevenson82} and
solve for the structure of a ``radiative zero atmosphere'' around a point mass
planet of mass $M_p$ and luminosity $L_p$. Since no energy is released in the
atmosphere, the luminosity is
\begin{equation}
L_p = -{16 \pi r^2\over 3} {\sigma_B\over \kappa\rho} {d T^4 \over dr} =
\hbox{ const}\;,
\label{Latm}
\end{equation}
where $\rho$ and $T$ are the gas density and temperature distance $r$
away from the protoplanet's centre. Further, writing the standard
equation of hydrostatic balance for the atmosphere, we find
\citep[see, e.g.,][]{Armitage10} for the temperature,
$T(r)$, and density, $\rho(R)$, profiles in the atmosphere:
\begin{equation}
T(r) = T_{\rm rad} = {GM_p \mu m_p\over 4 k_B r}\;,
\label{trad}
\end{equation}
\begin{equation}
\rho(r) = {64 \pi \sigma_B\over 3 \kappa_R L_p} T^4(r) r \quad \propto {1\over
  r^3}\;,
\label{rho_rad}
\end{equation}
where $\kappa_R$ is the Rosseland mean opacity and $\mu\approx 2.45$ is the
mean molecular weight.

This solution is approximately valid for $r\ge r_p$, where $r_p$ is the
planet's radius, and as long as the envelope's mass, $M_{\rm env} =
\int_{r_p}^{r_H} dr 4\pi r^2 \rho(r)$, is negligibly small compared with
$M_p$. Here $r_H = a (M_p/3M_*)^{1/3}$ is the Hill's radius of the planet
located a distance $a$ from the parent star of mass $M_*$.  The envelope's
mass is given by
\begin{equation}
M_{\rm env} = {\pi^2 \sigma_B \over 3 \kappa_R L_p} \left({GM_p \mu m_p\over k_B
  }\right)^4 \ln\left({r_H\over r_p}\right)\;.
\label{menv}
\end{equation}

We observe from equation \ref{menv} that at a constant $L_p$ the envelope's
mass is increasing as $M_{\rm env}\propto M_p^4$; therefore, there will be a
maximum $M_p$ at which the envelope's mass becomes comparable to $M_p$. As in
CA, the envelope in this case should contract rapidly due to its own
self-gravity. Thus, the condition $M_{\rm env} \approx M_p$ marks the critical
planet's mass above which a rapid accretion of additional mass from the
surrounding disc sets in.

Protoplanets are very bright at formation, up to $L_p\sim 10^{-2} \lsun$
\citep{Nayakshin10a}, and are extended, thus not much smaller than
$r_H$. Scaling the result to this fiducial value of $L_p$, and setting
$\ln(r_H/r_p)\sim 1$, we obtain
\begin{equation}
M_{\rm crit} \approx 6 M_J \; \left({L_p\over 0.01 \lsun}\right)^{1/3}
\left({\kappa_R\over 0.1}\right)^{1/3}
\label{mcrit}
\end{equation}

This derivation of the critical mass is rather approximate, as many additional
effects are important in real protoplanetary discs. For example, (a) the
protoplanet's luminosity decreases with time, (b) it is also a function of the
planet's mass itself; (c) the envelope's contraction luminosity may not be
neglected if its contraction becomes sufficiently fast; (d) the tidal field of
the star contributes at $r\simlt r_H$ to loosen the envelope; (e) the envelope
eventually may become optically thin, so that the temperature profile at large
$r$ would be shallower; (f) the envelope's temperature and density cannot fall
below those of the surrounding disc at large radii. Some of these effects act
to increase $M_c$, others to decrease its value. We could try to include some
of these effects in our analytical model here, but unfortunately this is not
worthwhile: the 1D approach cannot be very accurate in the given situation as
the flow of gas around the planet is distinctly a 3D pattern. Therefore, we
move on to 3D simulations in the next sections. In the next section we
describe a procedure with which we hope to capture the essence of the
radiative preheating effect in our planet-disc simulations.

\section{Numerical approach}\label{sec:numerics}

\subsection{Overall method}\label{sec:overall}

Our numerical approach follows that described in
\cite{ChaNayakshin11a} except for a different cooling function and
initial conditions. \cite{ChaNayakshin11a} used a cooling time
prescription that was a function of gas density only, smoothly joining
the low and the high gas density regimes. This treatment is best for
the gas within the protoplanet itself, as it follows the semi-analytic
model of molecular clump cooling due to dust opacity by
\cite{Nayakshin10a}. The relatively low density regime, as appropriate
for the ambient disc rather than the clump, was modelled with a simple
assumption of a constant cooling time of 100 years. This simple
approach yielded the appropriate radial range for disc fragmentation
as obtained both analytically and numerically, e.g., $R\sim 70$ AU
\citep[e.g.,][]{Rafikov05,BoleyEtal10}.

Here our focus is the material of the disc close to the planet's
location. Therefore we employ a more detailed although still approximate
approach to radiative cooling, following a number of authors
\citep[e.g.,][]{Johnson03,BoleyEtal10,VB10}. We write the energy loss rate per
unit time per unit volume as
\begin{equation}
\Lambda = \left( 36 \pi\right)^{1/3} {\sigma_B\over
  s} \left(T^4 - T_{\rm irr}^4 \right){\tau \over \tau^2 + 1}\;,
\label{lambda}
\end{equation}
where $T_{\rm irr}=10$ K is the background temperature set by external
irradiation, $\tau$ is the optical depth estimate, given by $\tau = \kappa(T)
\rho s$ \citep[cf.][]{CalvagniEtal12}. The lengthscale $s$ is given by $s =
({\cal M}/\rho)^{1/3}$, where ${\cal M} = 10 M_J$, and varies from a few AU to
tens of AU for the range of densities found in the simulation. The opacity law
is $\kappa(T) = 0.01 (T/10$K), as in the analytical model of
\cite{Nayakshin10a}.

As is well known, ``first cores'' hotter than $T \sim 2000$~K cannot
exist, since hydrogen molecules start to dissociate
\citep{Larson69}. This process presents a very large internal energy
sink, so that the clump behaves nearly isothermally at $T\simgt
2000$~K, which invariably leads to the second collapse during which
the first core contract to much higher densities and becomes the
``second core''. To mimic this behaviour, we impose a maximum
temperature for our gas, $T_{\rm max} = 2500$~K, and introduce a sink
particle inside the clump when the density of the clump exceeds
$\rho_{\rm sink} = 10^{-8}$ g~cm$^{-3}$. The sink radius is 5~AU.

We do not include the dust component in the present numerical simulations, in
contrast to \cite{ChaNayakshin11a}.

\subsection{Radiative preheating of the disc}\label{sec:rad_heat}

In our exploratory model we make a number of simplifying assumptions and
approximations. Firstly, the planet's luminosity is given by
\begin{equation}
L_p\left(M_p\right) = 1.5\times 10^{31} \;\hbox{ erg s}^{-1}\; \left({M_p\over
  10 M_J}\right)^{5/3} \;.
\label{lmodel}
\end{equation}
This is based on the analytical model of \cite{Nayakshin10a} at $t=0$. We note
that the contraction luminosity of a planet actually decreases with time, but
in our simulations the planet's mass may be evolving (usually increasing), and
therefore it is not clear what to take for $t=0$. This should not lead to
significant uncertainties in the results, however, because the planet's
luminosity at a constant $M_p$ decreases only by a factor of a few over the
duration of our simulations.

With this in mind, we calculate the radiative gas temperature, $T_{\rm rad}$,
a distance $r$ away from the centre of the clump assuming the radiative
equilibrium for the gas around the clump. We view $T_{\rm rad}$ as the {\em
  minimum} gas temperature at this location, arguing that any additional
heating processes not taken into account in this analytical treatment would
result in a higher temperature. For example, local contraction luminosity
(e.g., ``$P dV$'' work due to planet's envelope contraction) and shocks could
increase gas temperature above $T_{\rm rad}$. Operationally, we calculate the
gas temperature, $T$, at locations of SPH particles, solving the energy
equation that includes compressional heating and shocks minus the cooling rate
given by equation \ref{lambda}. If $T$ falls below $T_{\rm rad}$ we set $T =
T_{\rm rad}$.

When calculating $T_{\rm rad}$, we consider three different regions in
order of their distance $r$ from the clump's centre: (1) inside the
clump, $r \le r_p$; (2) outside the clump but within one disc
scaleheight, $H$, e.g., $r_p \le r \le H$; and finally, (3) $r> H$. As
gas density in the disc is largest in the disc midplane, decreasing in
directions perpendicular to it, most of the clump ``atmosphere'' mass
is located near the midplane of the disc. Therefore we now search for
an approximate temperature structure near the clump only along the
midplane of the disc.

The energy balance in the disc geometry dictates that the radiative
flux out of the optically thick disc in the direction perpendicular to
the midplane, $F_{\perp} \approx \sigma_B T^4/\tau$ is equal to the
column-integrated heating rate \citep[e.g.,][]{Shakura73}, where
$\tau$ is the optical depth estimated as described in section \S
\ref{sec:overall}. Since we are only setting here the minimum
temperature of the gas due to the radiative preheating from the
protoplanet, our radiative preheating energy balance requires that
$F_\perp \approx F_r$, where $F_r$ is the radiative flux from the
planet. In the region $r_p < r < H$, $F_r \approx L_p/4\pi r^2$, which
yields
\begin{equation}
T_{\rm rad} \approx \left[{ L_p (1+ \tau) \over 4\pi r^2
    \sigma_B}\right]^{1/4}\;,
\label{Trad1} 
\end{equation}
where we replaced $\tau$ with $(1+\tau)$ to account for the optically
thin limit in case $\tau \simlt 1$. While equation \ref{Trad1} appear
to be different from the radiative zero solution (equation \ref{trad})
for a spherical geometry, we shall see in section \ref{sec:atmosphere}
that the two profiles are actually quite similar in practice.

At distances $r \simgt H$ from the planet we
need to take additional consideration of the ``leakage'' of the radiative flux
from the disc in the vertical direction. An approximate solution formally
correct in the limit $H\ll R$ and a constant disc density (as a function of
$r$) yields that the radiation flux density at $r\gg H$ is reduced by the
factor $\sim \exp[-r/H]$. Thus we write for $r> H$
\begin{equation}
T_{\rm rad} \approx \left[{ L_p (1+ \tau)  \over 4\pi r^2
    \sigma_B} \exp \left( - {r\over H}+1 \right)\right]^{1/4}\;,
\label{Trad2} 
\end{equation}
which then makes $T_{\rm rad}$ continuous at $r=H$. 

Finally, inside the planet itself, $L_p$ cannot be considered
constant: the radiative flux is exactly zero at $r=0$ as $dT/dr = 0$
at the centre of the clump. Therefore we set $L_p(r) = (r/r_p)^2 L_p$
for $r\le r_p$ and then use this value of $L_p$ in equation
\ref{Trad1}. In practice this latter prescription makes the radiative
``preheating'' within the planet completely negligible, as of course
should be the case.

We also set $r_p = 5$~AU for simplicity when we calculate the clump mass ``on
the fly'' during simulations. Numerically, we find the densest part of the
protoplanet, define that location as the planet's centre, and then calculate
the current mass of the protoplanet, $M_p(t)$, as the mass within the fixed
radius $r_p$. We found that our planets -- that is dense gas clumps resulting
from gravitational collapse of the initial perturbations introduced at $t=0$
-- are always smaller than a few to 5 AU, and therefore we fixed $r_p =
5$~AU. This approach is conservative: if anything, it underestimates the
importance of preheating when the planet contracts to smaller sizes.  Finally,
we also fix $H=10$ AU when we calculate $T_{\rm rad}$ in equation \ref{Trad2}.

\subsection{Initial conditions and simulations performed}\label{sec:init}
%Initial condition: Disk_k0.01_T10_r200_Md325

The initial mass of the star is set to $M_*=0.7\msun$ in all the simulations.
The star is allowed to grow by accretion of gas from the protoplanetary disc
if the gas crosses the accretion radius of $10$~AU.

We are interested in a marginally gravitationally {\em stable}
protoplanetary disc that may self-regulate its mass by the episodic
creation of gas clumps that migrate inward quickly \citep{VB10}. At
the same time we would like to avoid non-linear clump-clump
interactions \citep[cf., e.g.,][]{BoleyEtal10,ChaNayakshin11a} as much
as possible, concentrating on one clump at a time. Therefore, we first
of all find a self-gravitating disc configuration that is on the verge
of fragmenting. In particular, we set a grid of simulations with discs
initially extending from 20 to 200 AU, with the radial profile
$\Sigma_0(R) \propto 1/R$, and the initial disc masses, $M_d$, in the
range from $0.2$ to $0.4\msun$. The disc material is initially set on
circular orbits around the protostar.

We then follow the disc evolution for $\sim 10$ orbits at the outer disc
edge. We found that the disc with the initial mass of $M_d = 0.325\msun$ does
not fragment, although shows 2 strong spiral arms. We note that the spiral
density waves in this rather high mass disc transfer angular momentum quickly,
as expected \citep{LodatoRice05}, so that the disc becomes less unstable as
time progresses. Therefore we are confident that the disc in this simulation
would not fragment even if we continued the simulation for an infinitely long
time.  On the other hand, the disc with an initial mass of $M_d = 0.35\msun$
fragmented on a number of clumps, with a complicated behaviour reminiscent of
that found in \cite{ChaNayakshin11a}. 

We thus select the disc with $M_d = 0.325$ at time $t\approx 10,000$
years as our starting initial condition. A density perturbation is
then inserted in the disc in a very simple way, e.g., by increasing
the gas density in a selected region by a factor of 4. In practice
this is done by adding new SPH particles in the region. These new
particles copy the density, temperature and velocity fields of the
existing SPH particles in the region; the coordinates of the new SPH
particles are shifted by a small distance, $h_s/5$, from their
``parent'' particles in random directions. Here $h_s$ is the SPH
smoothing length of the parent particle. Our results are essentially
independent of the exact procedure in which the perturbation is
inserted since we allow the region to contract and relax somewhat
before radiative preheating is turned on (see below).

\begin{figure}
\psfig{file=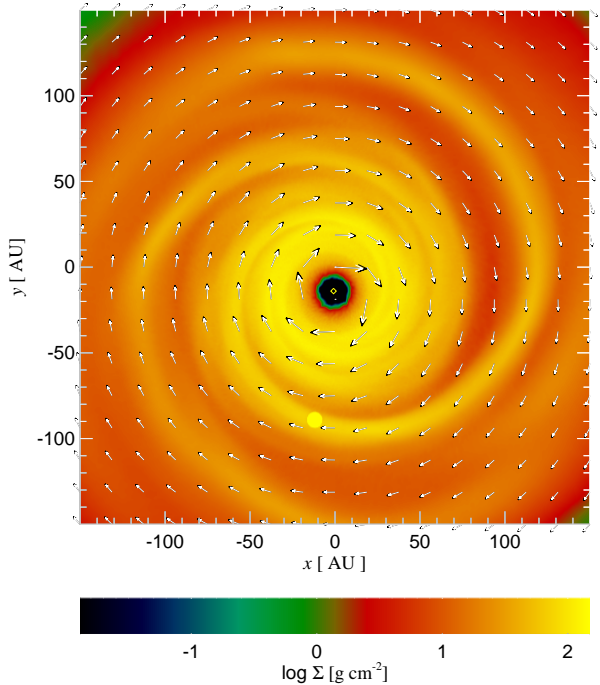,width=0.49\textwidth,angle=0}
\caption{The surface density of the initial marginally stable self-gravitating
  disc at time $t=0$, plus the initial perturbation (``protoplanet'') of mass
  $M_p = 4 M_J$, located at $R=75$~AU. The perturbation is inserted as
  described in \S \ref{sec:init}. The vectors in the figure show the gas
  velocity field.}
\label{fig:disc0}
\end{figure}

The location of the perturbation is set to be at the radial distance of 75 AU
from the parent star (which is close to the radial location of the minimum in
the Toomre's Q-parameter) at the densest part of the spiral arms. The surface
density of the initial disc with perturbation of mass $M_i = 4 M_J$ is shown
in Figure \ref{fig:disc0}. 

With this identical approach and numerical scheme we run a set of
simulations with and without planet feedback on its surroundings and
the initial perturbation mass varying between $0.5 M_J$ and $16 M_J$
is increments of factor of two. The corresponding runs are labelled MX
or MXF, where X is the mass of the initial perturbation in Jupiter
masses.

In the runs with feedback, the feedback is ``turned on'' after half a
rotation. This measure is taken to allow the initial perturbation to actually
contract gravitationally into a well defined self-bound clump before any
feedback is produced.

\begin{figure}
\psfig{file=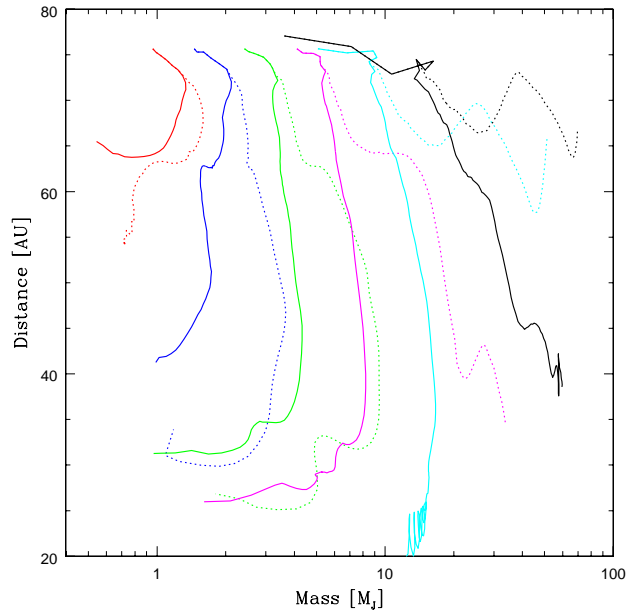,width=0.49\textwidth,angle=0}
\caption{Protoplanet-star separation versus planet's mass, $M_p$, for all of
  the simulations presented in this paper. The runs with feedback are shown
  with solid curves, whereas those without feedback are dotted. The
  protoplanet's initial masses are $0.5 M_J$ for the red curves, and increase
  by factor of two each time when moving to higher mass cases. See \S
  \ref{sec:overview} for more detail.}
\label{fig:RvsM}
\end{figure}

\section{Overview of Results}\label{sec:overview}

Figure \ref{fig:RvsM} presents the planet-star separation, $a$, of the
simulated planets versus their mass. The initial masses of the planets
  are $0.5 M_J$, 1, 2, 4, 8, and $16 M_J$ for the red, blue, green, violet,
  cyan and black curves, respectively. As explained at the end of section
\ref{sec:rad_heat}, to avoid ambiguities as to what is a planet, we measure
the planet's mass as that within a fixed radius of $5$~AU for this figure. We
find that at later times all the planets contract sufficiently to be within
$\sim 5$~AU. For the lowest mass perturbations, this procedure slightly
over-estimates the initial planetary mass somewhat, whereas for the highest
mass perturbation the planetary mass is underestimated (which is why the
  initial masses of planets in the figure may appear slightly different from
  the values quoted above). However, this convention affects only the first
$\sim$ half a rotation of the planets around their parent stars in the
simulations and has a very minor effect on the final results.

The end of the simulation here is defined either when the protoplanet is
completely destroyed by tidal torques, or gains a sufficient amount of mass
to become the ``second core'' \citep{Masunaga00}, which in these simulations
is always of a brown dwarf mass. 

In Figure \ref{fig:RvsM}, results for runs with feedback are shown by
solid curves, whereas those without feedback are dotted. Same colours
are used for runs with the same initial perturbation mass.  A planet
growing in mass at a constant separation would trace a horizontal
track, while a planet migrating radially at a constant mass would be
on a vertical line.

Initially, dotted and dashed curves coincide, so that planets migrate
inward and gain mass from their surroundings. This is simply because
the radiative feedback is switched on with a delay of half the initial
period to allow the clump to contract sufficiently. There is then a
bifurcation in trajectories of the planets once the feedback switches
on. Planets with feedback (solid curves) accrete gas at smaller rates
than their counterparts without feedback, as expected.  Relatively low
mass protoplanets, with $M_i=0.5, 1,$ and $2 M_J$ migrate inward more
rapidly than they contract independently of whether the feedback is on
or off. However, for higher mass protoplanets, starting with $M_i = 4
M_J$ (runs M4 and M4F), where the protoplanets end up in the Figure
depend drastically on the feedback. The $M_i=4 M_J$ planet without
feedback accretes mass very rapidly and becomes a brown-dwarf mass
object with mass of $M_p = 35 M_J$, having migrated to $a \approx
34$~AU. In contrast, the $M_i=4 M_J$ planet with feedback grows in
mass much slower, reaching the maximum mass of only $8 M_J$. At this
point the planet is already at $a\approx 30$ AU, where tidal torques
from the star start to unbind it. Eventually the planet migrates to
$a\sim 25$~AU and gets totally unbound.

Similarly, the $M_p = 8 M_J$ with feedback grows to $M_p = 17 M_J$;
but, having migrated too closely in, gets unbound at $a\sim
20$~AU. The same case without feedback becomes as massive as $M_p \sim
60 M_J$, at which point it collapses into the second
configuration. This proto-''planet'' migrates very little, ending up
close to where it started at the end of the simulation, and actually
evolving to $M_p \approx 130 M_J$ (see \S \ref{sec:M8}.

Finally, the heaviest perturbation runs considered here, $M_i = 16
M_J$, end up as sink particles, $M_p \approx 58 M_J$ and $M_p \approx
70 M_J$, for the feedback and no feedback runs, respectively. In both
of these cases the sink continues to grow in mass.  However, even in
this case (M16F), the feedback has apparently played a role in the
evolution of the protoplanet. The final separation of the sink
particle from the parent star is smaller in run M16F than it is in
M16.

To summarise, all of the runs presented here, with or without
feedback, yielded one of the two possible outcomes: (a) an inward
migration of lower mass clumps, followed by their tidal disruption at
the inner disc, or (b) rapid accretion of gas onto the more massive
protoplanets from the disc, followed by collapse to the second core
(proto brown dwarf). We shall see later on that the sink particles
introduced continue to grow rapidly. The end result of (b) is thus
formation of a low mass secondary star in orbit around the
primary. The role of radiative preheating from the protoplanet is to
only shift the dividing mass of the planet at which the transition
from evolutionary path (a) to that of (b) takes place.

\section{Clump accretion versus migration}\label{sec:M8}

\begin{figure}
\psfig{file=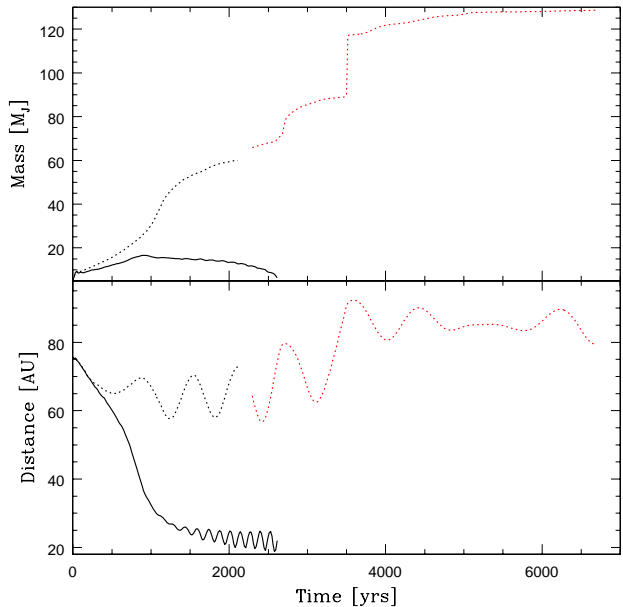,width=0.49\textwidth,angle=0}
\caption{Time evolution of the protoplanet's mass (top panel) and separation
  (bottom panel) for simulation M8 (dotted; no feedback) and M8F (solid; with
  protostellar feedback). In the simulation without feedback, the clump
  collapses into a second core, at which point we model it as a sink particle,
  and plot its evolution with the red color. This sink particle then continues
  to gain mass from the disc rapidly and grows to mass of almost $0.13 \msun$
  by the end of the simulation.}
\label{fig:vst}
\end{figure}

In this section we contrast the $M_i = 8 M_J$ simulation without
feedback (M8) with the same $M_i$ run but with feedback (M8F). These
particular simulations exemplify the effects of the radiative feedback
from protoplanets on its surroundings in the most clear way.

The top panel in Figure \ref{fig:vst} shows the time evolution of the
protoplanetary mass, whereas the bottom panel of the figure shows the
planet-star separation versus time. The solid and dotted curves correspond to
simulations with feedback and without, respectively. The dotted curve has a
discontinuity in it to emphasise the ``phase transition'' that occurs in the
evolution of simulation M8. Before time $t\approx 2300$~yrs, the clump is
molecular, e.g., it is the first core in the usual terminology
\citep[e.g.,][]{Larson69}, whereas after that time molecular hydrogen in the
clump dissociates, so that the clump collapses to much higher densities and
becomes the second core. The latter phase of the evolution is shown with
the red color. We cannot continue to resolve such high densities due to
numerical limitations. A sink particle with sink radius of 5~AU is introduced
at that point. 

The sink continues to accrete gas from the disc. Note that its accretion rate
is not steady. There is a particularly rapid accretion episode at $t\approx
3500$~yrs, when the secondary sink disrupts and then accretes a gas clump that
migrated from the outer disc. The latter clump was not introduced in the disc
artificially; instead it formed during the simulation due to the perturbations
of the ``original'' protoplanet we inserted. More on this below in this section.

In contrast, in simulation M8F, the mass of the protoplanet varies by about a
factor of two while it spirals in on the timescale of about $10^3$ years. The
simulation ends with the planet destroyed by tidal forces at $a\sim 20$~AU.

\begin{figure*}
\centerline{\psfig{file=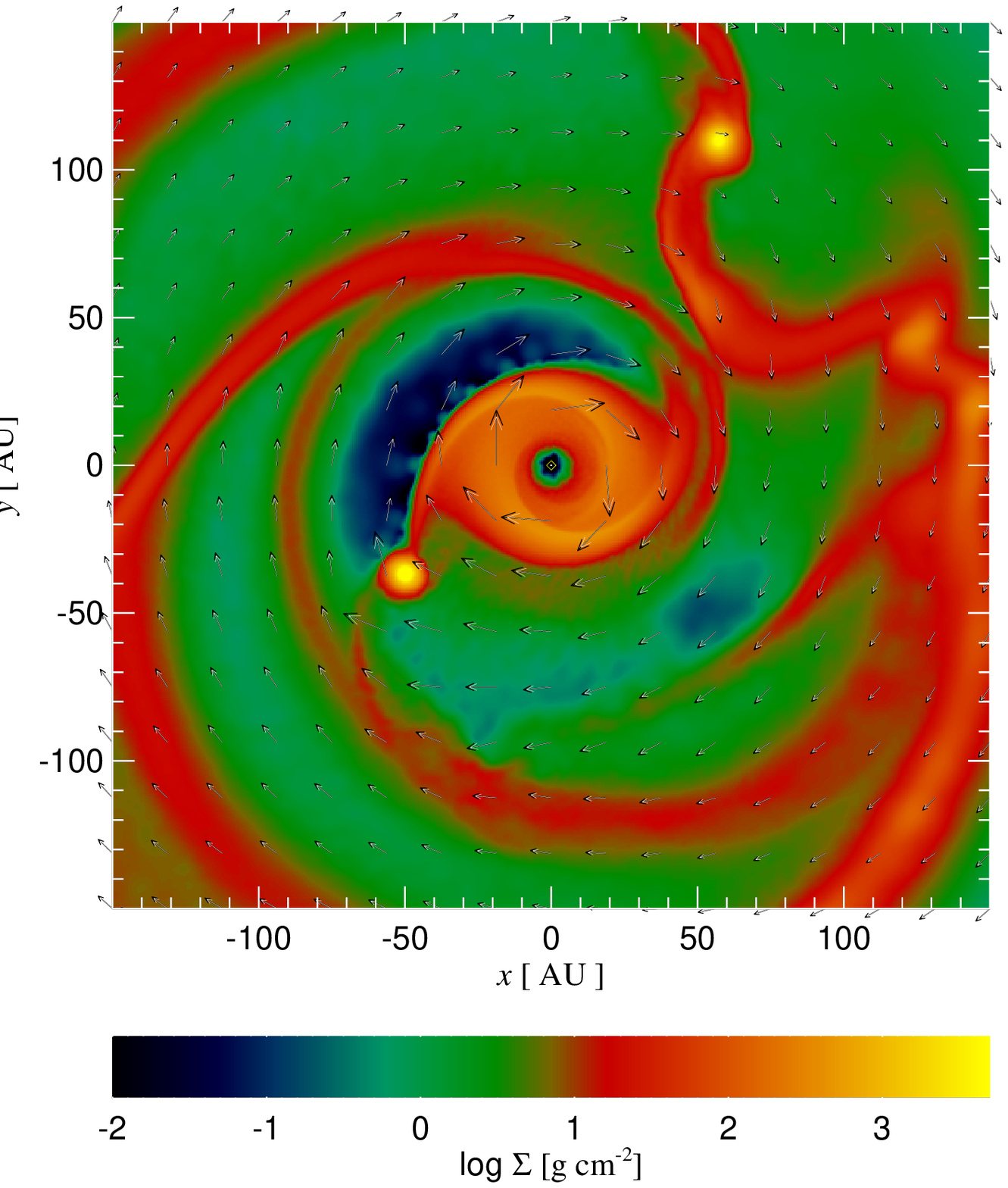,width=0.5\textwidth,angle=0}
\psfig{file=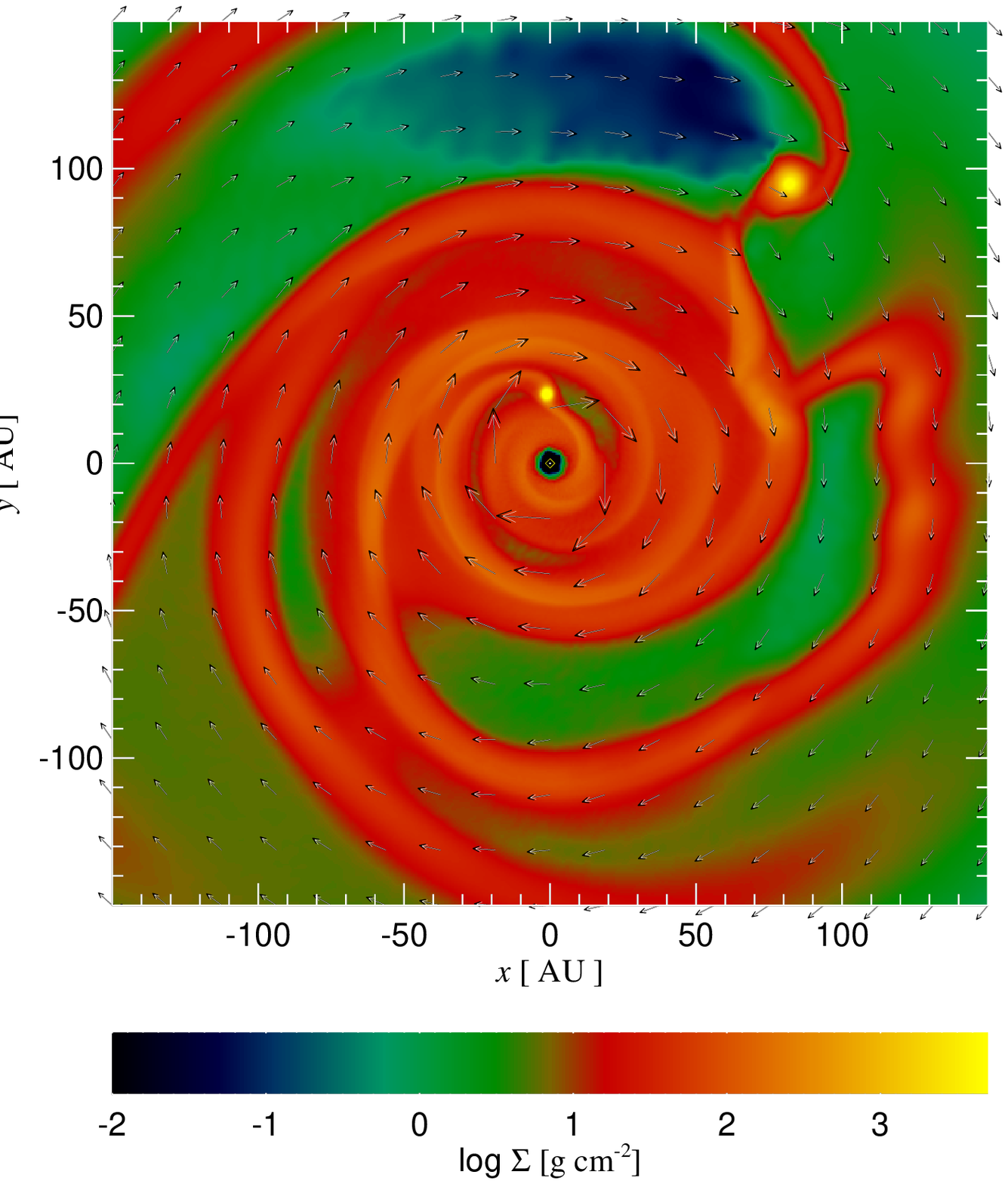,width=0.5\textwidth,angle=0}}
\centerline{\psfig{file=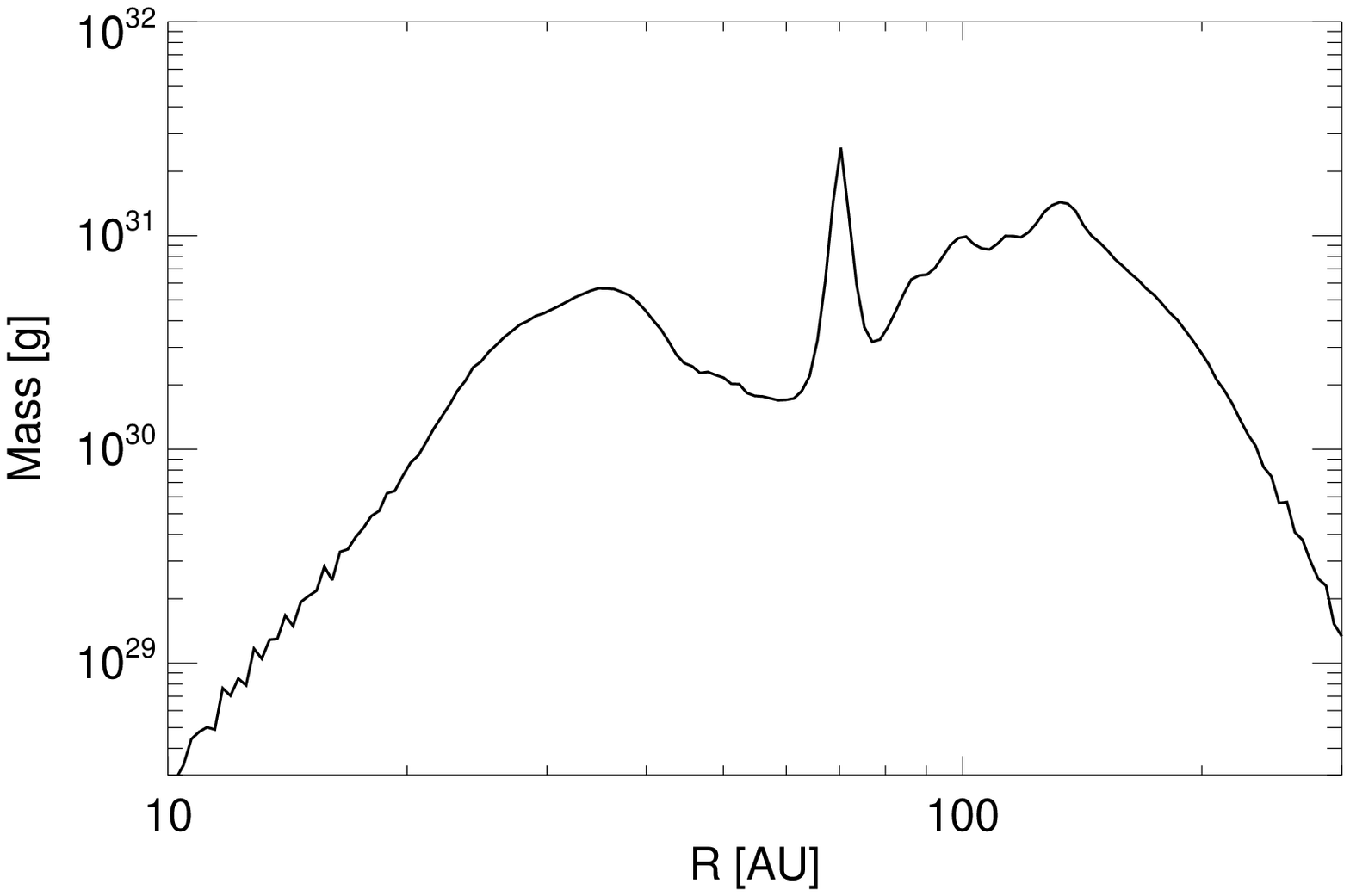,width=0.5\textwidth,angle=0}
\psfig{file=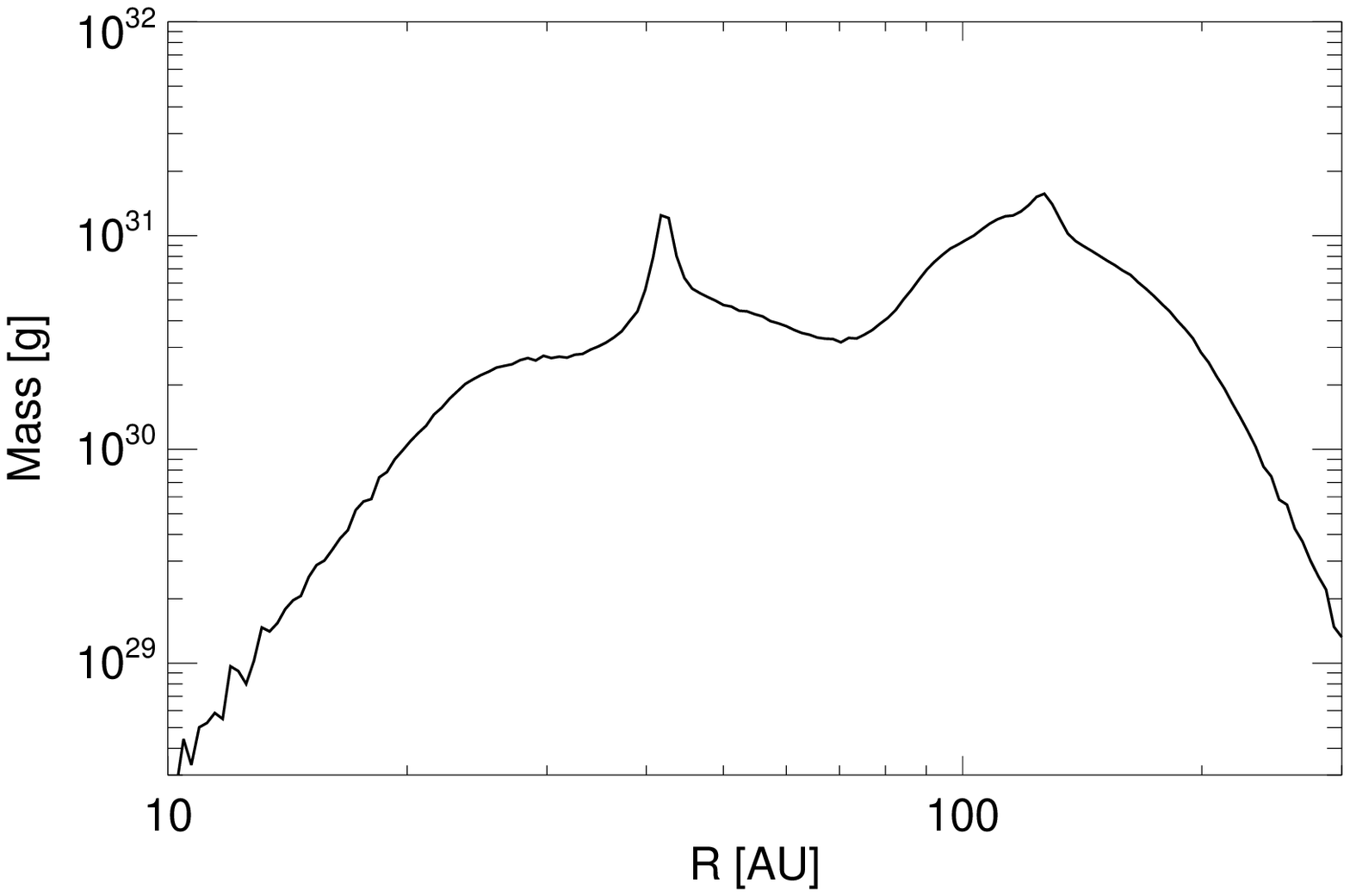,width=0.5\textwidth,angle=0}}
\centerline{\psfig{file=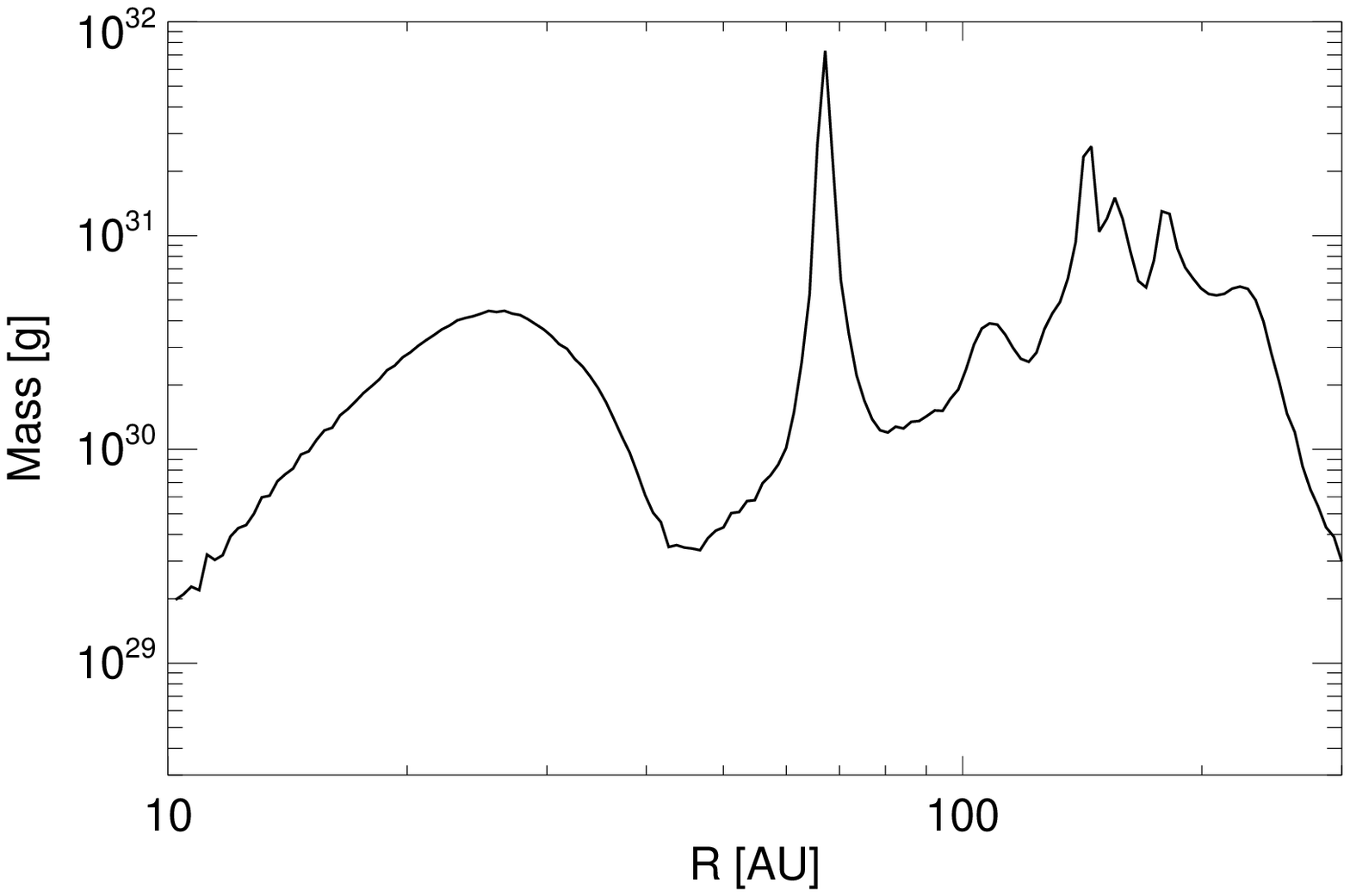,width=0.5\textwidth,angle=0}
\psfig{file=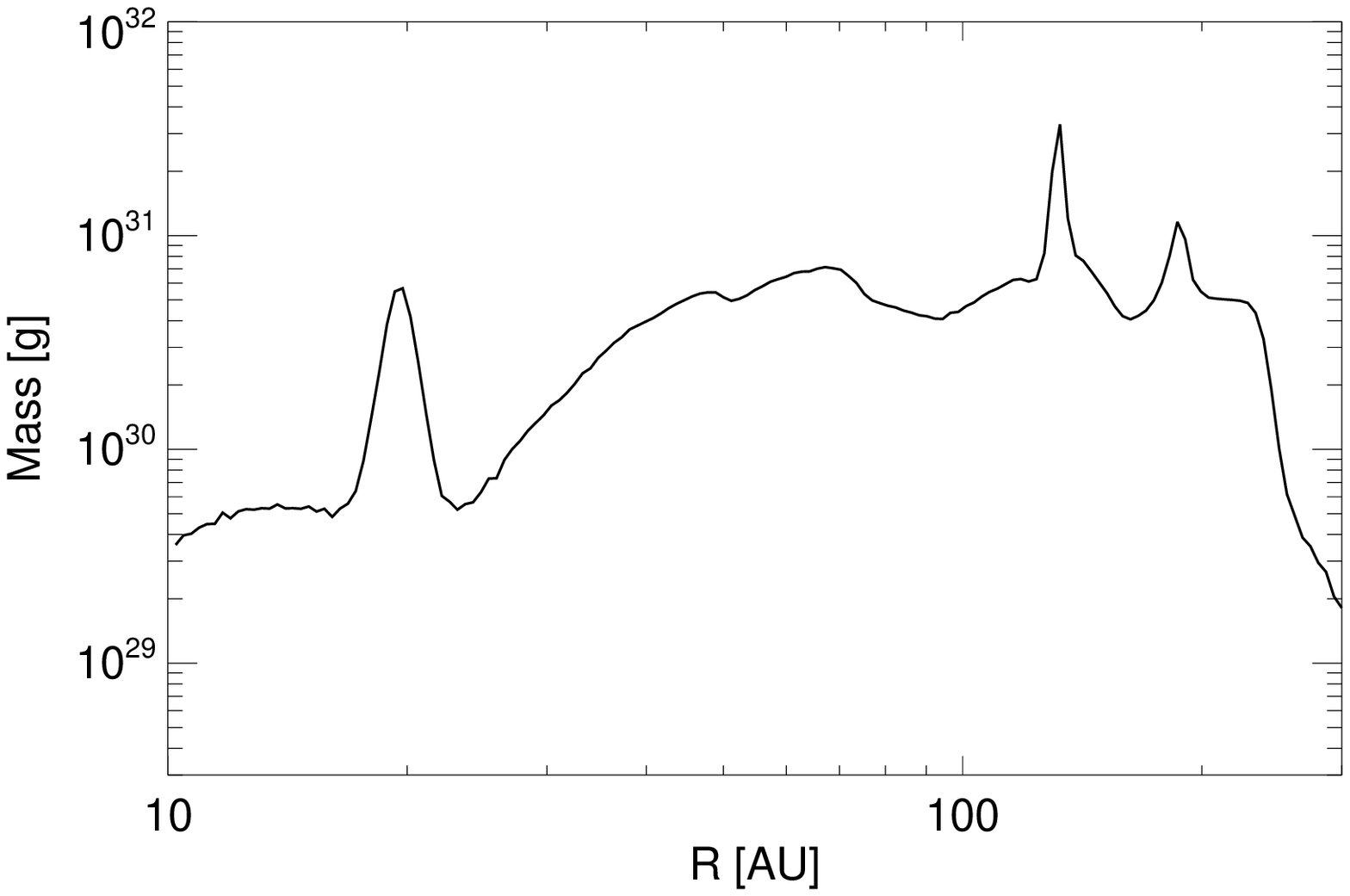,width=0.5\textwidth,angle=0}}
\caption{{\bf Left column}: simulation M8 (no feedback), {\bf Right}:
  simulation M8F, with feedback. The top row of panels show the disc
  surface density profiles at time $t=2000$~yrs, whereas the row in
  the middle and on the bottom show gas mass in radial bins at time
  $t= 1200$ and 2100 yrs, respectively. See \S \ref{sec:M8} for more
  detail.}
\label{fig:disc_vs_clump}
\end{figure*}

Figure \ref{fig:disc_vs_clump} presents the same simulations but
focuses on the interaction between the planet and the disc. In figure
\ref{fig:disc_vs_clump}, left and right column panels correspond to
simulations M8 and M8F, respectively. The top most panel in the figure
show the gas column density at time $t=$ 2000~yrs for the
simulations. In both cases the gas clumps at the top right corner, and
further gas concentrations on the way to contracting into clumps at
$R\sim 130$ AU, are not the ones we are interested in. As explained in
section \ref{sec:init}, we avoided as much as possible formation of
these additional clumps by using the initial condition with a
marginally stable disc. We emphasise that no clumps develop in the
control simulation that we used to initialise the disc (cf. section \S
\ref{sec:init}). Therefore, it is clear that these ``uninteresting''
clumps were spawned by gravitational disc instability {\em amplified}
by the introduction of the clump that we do study here.  We do not
introduce feedback inside and around the second generation clumps as
we are not interested in those.

The first generation clumps are located in the inner disc in both cases. In
the case without feedback, the clump stalls close to its initial location, at
$R\sim 75$ AU. In the simulation with feedback, on the other hand, the clump
is at $R\sim 25$ AU at that time (these results are in accord with figure
\ref{fig:vst}, of course).

The middle and the bottom row of panels in figure
\ref{fig:disc_vs_clump} show the total gas mass in radial annuli (for
information, the bin size is proportional to radius $R$) in the
simulation M8 and M8F (left and right, respectively). The protoplanet
is identifiable as a strong maximum inside 100 AU; additional maxima
outside this region are the secondary clumps that we do not study
here.

The effects of the protoplanet in run M8F on its surrounding disc are
barely discernable, whereas in the no-feedback case M8 the protoplanet
strongly alters the disc structure. Namely, in the latter case the
protoplanet almost succeeds in carving out a gap around its radial
position. The deficit of gas near the planet's location in simulation
M8 is visible in the top (left) panel of the figure, and is further
quantified in the middle and the bottom left panels. There is a deep
depression in the disc mass on both sides of the planet. It is clear
that this protoplanet settles into the type II migration regime, and
this must be the main reason why the planet's migration rate is much
slower than that for the other simulation.

In the simulation with feedback, on the other hand, the protoplanet's
mass stays low enough to not open the gap in the disc. Indeed, the
panels in the right column of figure \ref{fig:disc_vs_clump} show that
there is not even a depression in the radial profile of the disc
mass. As argued by \cite{BaruteauEtal11}, when no gap is opened,
protoplanets are expected to migrate inward very rapidly, i.e., nearly
dynamically. In our simulation this situation persists until the
planet arrives near the inner edge of the disc. Note that the planet's
orbit is then not exactly circular and the eccentricity increases with
time (cf. the bottom panel in figure \ref{fig:vst}).

The main driver of these differences in the planet-disc interaction is
the mass of the protoplanet, whose evolution is strongly influenced by
the radiative feedback (preheating) from the clump, as we now show.

\section{The atmosphere of the clump}\label{sec:Atmosphere}

We now consider the behaviour of the material near the protoplanet in
simulations M8 and M8F. As explained previously, these two cases differ
significantly at late times, when both the radial position and the mass of the
protoplanets diverge (e.g., see fig. \ref{fig:vst}). For a meaningful
quantitative comparison we therefore contrast the clumps soon after the
feedback is turned on, just as their evolution starts to take different paths,
at time $t=500$~years (cf. figure \ref{fig:vst}). Both the mass and the radial
position of the clumps are approximately the same at that time.

\begin{figure*}
\psfig{file=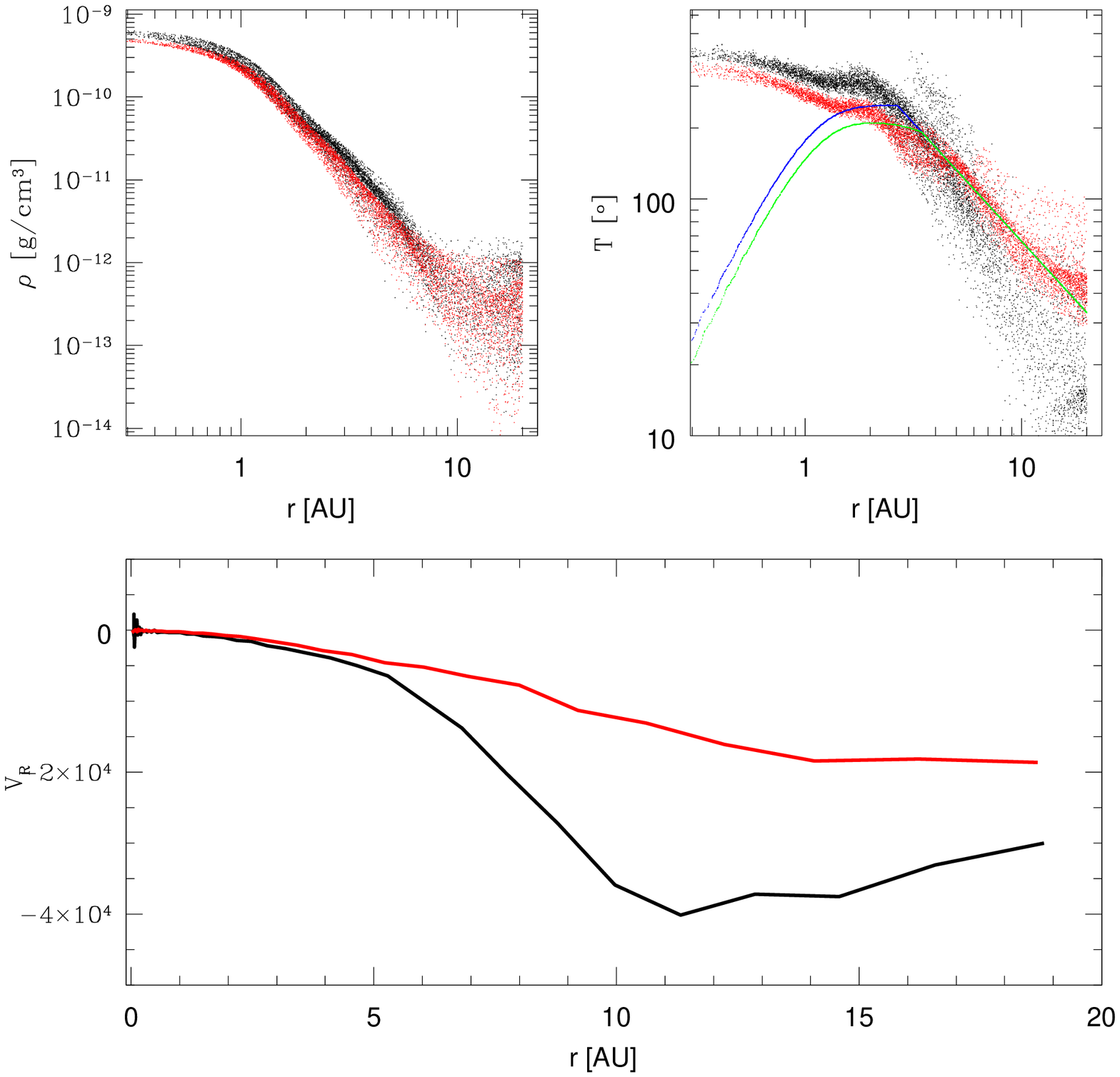,width=0.85\textwidth,angle=0}
\caption{Internal structure of two clumps for a simulation M8F with radiative
  feedback (red color) and M8 (without feedback, black color) at time
  $t=500$~yrs. Top left and top right panels show SPH particle temperature and
  density, respectively. The blue and the green curves show the radiative
  temperature $T_{\rm rad}$ from equation \ref{Trad1}.}
\label{fig:clump_structure}
\end{figure*}

Figure \ref{fig:clump_structure} shows temperature, density and radial
velocity of gas particles as a function of distance, $r$, from the
protoplanet's centre (defined as its densest point). Red and black
dots and curves correspond to simulation M8F and M8, respectively. In
the top panels dots show temperature (left) and density (right panel)
at the location of individual SPH particles. The left top panel shows
that the density in the atmosphere of the protoplanet without feedback
is $\sim 2-3$ times higher than in simulation M8F.

The curves in the right panel of figure \ref{fig:clump_structure} show
the radiative zero solution temperature, $T_{\rm rad}$, defined by
equation \ref{trad}. Here $M_p(r)$ is the minimum of the enclosed mass
within radius $r$ and $8 M_J$. These curves play no role in the actual
simulation, but are presented here to make a comparison with the
radiative zero solution.

There is a significant dispersion in SPH particles' temperature at the
same distance from the planet's centre, which is to be expected given
that the geometry ceases to be quasi-spherical and becomes disc-like
farther away from the planet. Nevertheless, it is clear that the gas
around the planet at $r\sim 10$~AU is significantly colder than
$T_{\rm rad}$ for the simulation without feedback (M8). At the same
time, gas may be significantly hotter at the same location for
simulation M8F, with the mean value of $T$ following $T_{\rm rad}$
approximately.

The importance of this difference in the temperature profile near the
planet is best appreciated in the bottom panel of figure
\ref{fig:clump_structure}, where we show the mean radial velocity
(defined with respect to the planet's centre, of course) of SPH
particles for the two simulations. The simulation without feedback
(black curve) has a significant negative radial velocity, indicating
infall, e.g., accretion. The maximum infall velocity is nearly 0.4
km~s$^{-1}$ at $\sim 10$~AU, which is slightly above the isothermal
sound speed for molecular gas at temperature of 30 K, $c_s = 0.3$
km~s$^{-1}$ (cf. the black dots at this distance in the top right
panel). This shows that gas is contracting quite rapidly onto the
protoplanet in simulation M8, fuelling its rapid growth in mass.

In contrast, in simulation with radiative preheating (M8F, red curve),
the maximum infall velocity, reached at $\sim 15$~AU, is a fraction of
the gas sound speed, and becomes several times smaller than that
closer in. Gas accretion is thus strongly hindered due to larger
pressure gradient around the clump, which in itself is due to
radiative preheating of the material near planet. This protoplanet is
thus surrounded by a quasi-static atmosphere, as expected based on our
approximate analytical theory.

\begin{figure*}
\psfig{file=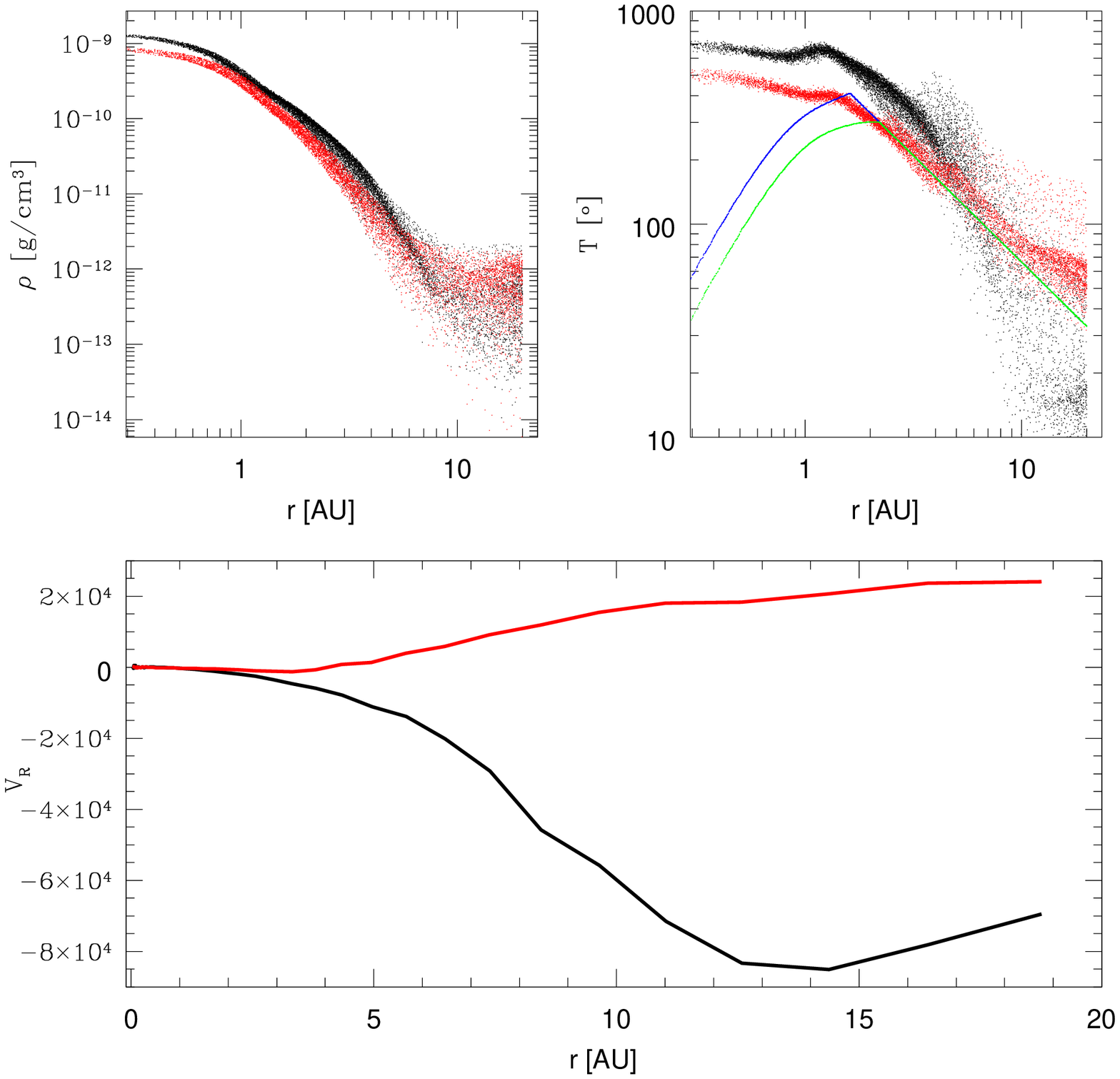,width=0.85\textwidth,angle=0}
\caption{Same as figure \ref{fig:clump_structure} but at time
  $t=900$~yrs. Note that the radiative preheating case M8F develops a positive
  radial velocity at that time (red curve in the bottom of the figure).}
\label{fig:clump_structure2}
\end{figure*}

These differences only get amplified at later times, as Figure
\ref{fig:clump_structure2} shows, where we show exactly same quantities but at
time $t=900$~yrs. The M8F clump actually starts to loose mass rather than gain
it (cf. fig. \ref{fig:vst}), as is seen from a positive radial velocity in the
bottom panel. This is in stark contrast to simulation M8 (no feedback), where
the inward gas velocity reaches almost 1 km~sec$^{-1}$.

\section{Discussion}\label{sec:discussion}

In this paper we studied the evolution of a self-gravitating gas clump in a
marginally gravitationally stable disc. We found that below a certain critical
clump mass, $M_{\rm cr}$, gas clumps migrate inward and get tidally disrupted
quickly. Above this mass the clumps accrete gas from the disc rapidly,
eventually becoming as massive as several tens of Jupiter masses. At such a
high mass the gas clumps collapse into the second core configuration due to
Hydrogen molecules dissociation \citep{Larson69,Masunaga00}.  These clumps
become too massive for the disc to be transported inward rapidly. Therefore
they stall and continue to grow by accretion, becoming massive brown dwarfs or
low mass companions to the host star.

We also studied how this behaviour changes if the radiative output of the
protoplanet due to its contraction luminosity is taken into account. We found
that the qualitative behaviour of the simulations did not change, except the
value of $M_{\rm cr}$ increases. In particular, in the simulations without
radiative preheating from the planet, we obtained $M_{\rm cr}$ between 2 and 4
$M_J$, whereas with preheating $M_{\rm crit}$ rises to between $8 M_J$ and $16
M_J$. This result is consistent with the approximate analytical argument,
presented in \S \ref{sec:analytical}, similar in spirit to the well known Core
Accretion result \citep{Mizuno80,Stevenson82}. This argument shows that
radiative preheating from the clump may be sufficiently important to arrest
gas accretion for clumps smaller than $\sim 6 M_J$ (see equation
\ref{mcrit}). Having said this, we note that for clumps less massive than
  $\sim 2 M_J$, inward migration appears so rapid that the outcome is similar
  whether the pre-heating is included or not: the clump is tidally disrupted
  at a few tens of AU.

The qualitative behaviour found in this paper -- migration and tidal
disruption of low mass clumps, but migration stalling and rapid gas
accretion commencing for high mass clumps -- is likely to be general
for marginally stable discs. We are however certain that the critical
mass $M_{\rm cr}$ itself is likely to depend on the disc opacity,
mass, radial location of where the gas clump is born, and the
surrounding temperature of the disc. For example, higher disc
temperatures may render radiative preheating from the clump
ineffective in supporting the radiative atmosphere. A fuller
investigation of the parameter space is planned in the near future.

Despite the fact that our study only covers a small fraction of the
available parameter space, we believe that the sensitivity of the
final fate of the clump to its initial mass (and other parameters in
the problem, as discussed above), and such detail as radiative
preheating of gas near the clump, is an important result to note. We
believe this sensitivity may be the reason why different groups of
numerical simulators obtain widely differing results, with some
finding that massive self-gravitating discs around young stars mainly
fragment onto brown dwarfs and low mass stars \citep{SW08,SW09b},
whereas others find that young gaseous clumps mainly migrate and get
tidally disrupted \citep[e.g.,][]{BoleyEtal10,ChaNayakshin11a}, see
also \cite{ZhuEtal12a}. In our opinion, it is possible that relatively
small differences in the numerical techniques and/or regions of the
parameter space studied may be sufficient to lead to divergent results
obtained in those previous studies.

\section{Acknowledgments}

Theoretical astrophysics research in Leicester is supported by an STFC Rolling
Grant. This research used the ALICE High Performance Computing Facility at the
University of Leicester, and the DiRAC Facility jointly funded by STFC and the
Large Facilities Capital Fund of BIS.

%\bibliographystyle{mnras}
%\bibliography{../nayakshin}

\label{lastpage}

\end{document}